# Synthesis of standard 12-lead electrocardiograms using two-dimensional generative adversarial networks

Yu-He Zhang, PhD,[*] Saeed Babaeizadeh, PhD

*Advanced Algorithm Research Center, Philips Healthcare, Cambridge, MA, USA*

**Abstract**

This paper proposes a two-dimensional (2D) bidirectional long short-term memory generative adversarial network (GAN) to produce synthetic standard 12-lead ECGs corresponding to four types of signals—left ventricular hypertrophy (LVH), left branch bundle block (LBBB), acute myocardial infarction (ACUTMI), and Normal. It uses a fully automatic end-to-end process to generate and verify the synthetic ECGs that does not require any visual inspection. The proposed model is able to produce synthetic standard 12-lead ECG signals with success rates of 98% for LVH, 93% for LBBB, 79% for ACUTMI, and 59% for Normal. Statistical evaluation of the data confirms that the synthetic ECGs are not biased towards or overfitted to the training ECGs, and span a wide range of morphological features. This study demonstrates that it is feasible to use a 2D GAN to produce standard 12-lead ECGs suitable to augment artificially a diverse database of real ECGs, thus providing a possible solution to the demand for extensive ECG datasets.

*Keywords:*   12-lead ECG, data synthesis, electrocardiogram, generative adversarial network (GAN)

## Introduction

The design of algorithms to diagnose cardiac conditions based on automatic analysis of multi-lead electrocardiogram (ECG) recordings requires access to a comprehensive ECG database. It is challenging to ensure that such a collection captures the variety of real-world ECGs observed in clinical practice. This is particularly true in the context of rarely observed but clinically significant signals. Moreover, substantial effort is required to annotate the recordings accurately to enable their use as gold standards during both the training and the evaluation of an algorithm. Therefore, the production of synthetic ECGs with known annotations to augment clinically acquired databases is essential.

Since their introduction in 2014, generative adversarial networks (GANs) have enjoyed continued success in tasks related to the production of synthetic images, including high resolution images of human faces. The level of fidelity attained is such that even image experts have difficulty discerning whether the resulting images are synthetic or natural [1], [2]. The GAN architecture has also been successfully applied in areas other than the spatial domain during the six-year period since its invention.

Time series are among the simplest data structures used to represent series of non-independent temporal events. Recurrent neural networks (RNNs) have been widely used to address the "memory" of time series (i.e., the dependence of current events on past events). Among the numerous RNN variants, a combination of long short-term memory (LSTM) and convolutional neural network (CNN) has achieved noticeable success in generating coherent time series [3], [4]. Various types of time series have been explored, including stock-market predictions, audio sequences [5], and records of patient monitoring data [6].

ECGs are among the most complicated time-series data appearing in medical records. They provide rich and comprehensive information on patients' cardiovascular conditions and their dynamics. High quality stratification of any cardiovascular condition requires extensive ECG datasets. Significant effort has been expended to construct ECG databases that are publicly available to clinicians and medical researchers. Nevertheless, good quality ECG databases suitable for research purposes remain difficult to find. Given the recent advancement of deep learning-based methods, which require extensive training sets to guarantee high-quality learning, such databases have become even more necessary, especially for low-prevalence diseases. Many researchers use proprietary databases, but this can lead to data-privacy conflicts. Synthesis of medical time series—ECGs in particular—is a promising alternative [7]–[14]. High-fidelity synthetic ECGs could be used to augment existing databases, thereby facilitating training in deep-learning-based methods.

    * Advanced Algorithm Research Center, Philips Healthcare, Cambridge, MA 02141, USA.
    *E-mail address:* yu-he.zhang@philips.com

Time series are one-dimensional in nature. Most existing approaches to ECG synthesis based on GAN/RNN generate ECG lead signals that are independent of each other. They are either trained on single-lead ECGs, or extracted from multi-lead ECGs but trained independently of the leads. The metric of evaluation of such methods is also not well defined, and usually depends on simple visual inspection—the learning process is considered to be successful if the generated time-series data resemble ECG signals (with QRS complexes or occasional distinguishable T-waves and P-waves). The diagnostic capabilities of single-lead ECGs are limited. The accurate diagnosis of most cardiovascular conditions requires multiple ECG leads to capture a sufficient amount of information. Twelve-lead ECGs are used in sophisticated ECG devices, but, owing to their cost and operational complexity, they are not prevalent outside of healthcare settings. Efforts have been made to reconstruct standard 12-lead ECGs from more economical configurations (usually with 1–3 leads) [15]–[20]. The goal of these methods is to convert a single-lead, 2-lead, or 3-lead ECG to a 12-lead ECG instead of synthesizing new 12-lead ECGs. To the best of our knowledge, there is no published study on generating synthetic standard 12-lead ECGs using deep learning.

Deep-learning-based ECG-generation models face challenges in the following areas:

a) generating consistent multi-lead (12-lead in particular) ECGs that are physiologically plausible;
b) verifying whether the generated ECGs mimic the intended clinical conditions;
c) defining an automated metric of evaluation that does not depend on human visual inspection; and
d) generating multiple ECGs of the intended type that are also sufficiently distinct from each other under the constraint of feasible resource allocation.

This study aims to address all of the aforementioned challenges. To this end, we propose a two-dimensional bidirectional LSTM GAN (2D BiLSTM GAN) model to synthesize 12-lead ECGs. The second dimension is meant to capture the physiological and spatial correlation among ECG leads of the same recording. The production and verification of synthetic 12-lead ECGs are first described in detail. Then, the results obtained via the proposed synthesis process are presented. Finally, we present and evaluate the diversity of the synthetic ECGs and investigate the existence of any bias towards the training set that could induce concerns of overfitting of the model.

## Production of Synthetic 12-Lead ECGs

As depicted in Fig. 1, the end-to-end process to produce synthetic ECGs consists of the following stages:

1. Data preprocessing;

2. Setup and training of the GAN model;
3. Post-learning data processing;
4. Verification of the synthetic ECGs.

Each of these stages is described in detail in the following subsections.

### Data Preprocessing

The ECG waveforms used for this study were extracted from multiple databases in the public and private domains. The public databases included PTB-XL [21], CCDD [22], CSE [23], and Chapman [24]. The private ECGs were from multiple databases and had been recorded in a variety of clinical settings (emergency departments, chest-pain centers, clinics, etc.) using commercial cardiographs. Regardless of the database of origin and any annotations, the Philips DXL™ ECG algorithm [25]-[34] was used to re-classify the ECGs to build the training databases for the four ECG categories that will be discussed later in this section. In the final stage of the synthesis process, the DXL™ algorithm was used again to verify that matching morphological profiles could be generated.

It should be emphasized that, although various ECG classifiers have been studied [35] and the advantage of expert manual validation has also been evaluated [36], using the DXL™ as the sole classifier was not meant to provide gold standard databases or rules. It was mainly for the consistent and coherent classification rules being applied throughout this study for real ECGs used for the model training and for the synthetic ECGs for the final verification and evaluation. The DXL™ is a diagnostic ECG algorithm that provides an analysis of the amplitudes, durations, and morphologies of ECG waveforms. ECG waveform analysis is based on standard criteria for interpretation of these parameters, calculations of the electrical axis, and the relationship between leads. The DXL™ algorithm uses the mean representative beat process to provide the cleanest possible measurements. The mean value is derived by aligning the complexes; comparing all beats in a 10-second recording against each other and selecting only the similar beats; and then taking the average at each sample point. Stringing these points together creates an averaged representative beat. Raw 12-lead 10-second recordings were automatically analyzed via the DXL™ algorithm to classify them and generate 12 representative beats corresponding to each recording—one representative beat per lead. Each representative beat of each lead was taken to be an ECG shape averaged over 1,200 milliseconds at a sampling rate of 1000 samples per second (SPS). Then, an 800-millisecond segment centered on the QRS complex was selected from the representative beat, and it was further down-sampled to 500 SPS.

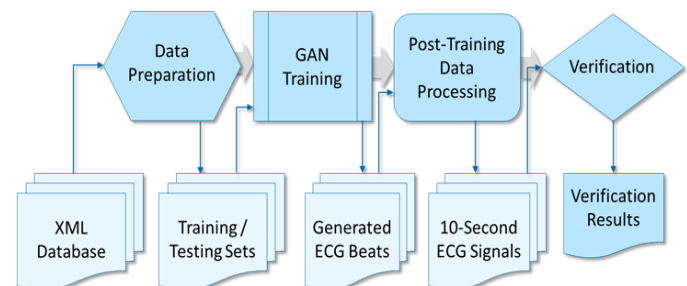

Fig. 1. End-to-end process to produce synthetic 12-lead ECGs via a generative adversarial network (GAN).

To aid the efficient learning of meaningful features by the GAN without overwhelming and possibly confusing it with irrelevant features, a number of unnecessary variables were removed in a stepwise fashion from the training dataset. These preprocessing steps considered the limitations of available model architectures, the size of the synthetic database that could reasonably be constructed from the available real-patient data, and the computational resources of the system used during training.

First, the training datasets were divided into different categories based on the primary clinical conditions they represented, such as Normal or left ventricular hypertrophy (LVH). In this study, we focused on four categories—Normal, LVH, Left Bundle Branch Block (LBBB), and Acute Myocardial Infarction (ACUTMI). In each given training cycle, the GAN model was only trained in one category. For instance, only the Normal ECG dataset was used to train a model that can generate synthetic Normal ECGs.

Second, by excluding ECGs of patients with heart rates higher than 100 beat per minute (BPM), it was ensured that the 800-millisecond representative beat section contained only one full ECG cycle, including the P-wave and the T-wave. It should be noted that for ECGs with significant prolonged QT intervals, the end of T-wave may not be included in the 800-millisecond window. This phenomenon might have affected a limited number of ECGs, but certainly did not have any material impact on the validity of the technique.

Third, the QRS complex, which is the predominant feature of an ECG and among the first ones learned by the GAN, was selected to be the anchor of each beat cycle. Therefore, all ECG beats prepared for training were centered at the QRS complex, with the center being determined by $Qo + QRS_{dur}/2$, where $Qo$ is Q wave onset and $QRS_{dur}$ is the duration of the QRS complex.

Fourth, as only two leads out of six limb leads in a 12-lead ECG are independent, only leads I and II were used during training. Following the generation of synthetic ECGs, the other four limb leads (III, aVR, aVL, and aVF) were derived using the following simple formula to reconstruct the final 12-lead ECGs:

$$\begin{bmatrix} III \\ aVR \\ aVL \\ aVF \end{bmatrix} = \begin{bmatrix} -1 & +1 \\ -\frac{1}{2} & -\frac{1}{2} \\ +1 & -\frac{1}{2} \\ -\frac{1}{2} & +1 \end{bmatrix} \begin{bmatrix} I \\ II \end{bmatrix}. \tag{1}$$

The sizes of the training sets corresponding to the four ECG categories are listed in Table 1. The databases were split between training and testing as a way to accommodate the demanding dataset size the GAN model needed to learn the morphological features in order to generate converged outputs. The size of the testing dataset was sufficient for statistical evaluation.

Table 1
Training Set Sizes for Each ECG Category

| ECG Category | Total | Training | Testing |
|---|---|---|---|
| Normal | 10,013 | 9,011 | 1,002 |
| LVH | 11,519 | 10,367 | 1,152 |
| LBBB | 10,080 | 9,072 | 1,008 |
| ACUTMI | 11,214 | 10,092 | 1,122 |

The distribution of ECGs from public and private databases for each category is listed in Table 2.

Table 2
The Number of ECGs from Public and Private Databases for Each Category

| ECG Category | PTB-XL | CCDD | CSE | Chapman | Private |
|---|---|---|---|---|---|
| Normal | 3,501 | 0 | 0 | 1,429 | 5,083 |
| LVH | 0 | 0 | 0 | 0 | 11,519 |
| LBBB | 471 | 932 | 781 | 41 | 7,856 |
| ACUTMI | 0 | 5,501 | 0 | 0 | 5,713 |

All the ECGs were from adult patients. Patient age and sex may be considered when the DXL™ algorithm diagnoses and classifies cardiac conditions. Among the four ECG categories in this study, LBBB is treated by the DXL™ as age and sex agnostic. The impact of age and sex on classifying LVH was also statistically insignificant except that patients under 35 will not be considered for LVH. On the other hand, the criteria for diagnosis of ACUTMI vary by age group and sex. For instance, the DXL™ does not look for ACUTMI in males under 20 and females under 30. The ST elevation thresholds for ACUTMI are generally 100 uV for all leads except V2 and V3. For V2 and V3, the DXL™ uses ST elevation > 150 uV for women of any age, ST elevation > 200 uV for men 40 and older, and ST elevation > 250 uV for men under 40. Normal ECGs are the results of ruling out all possible abnormalities and are consequently dependent on age and sex. Ideally, for training and synthesizing ECGs, the ACUTMI and Normal databases should be segregated by sex and age groups. Nevertheless, with the ECG records that we were able to collect, the databases with more refined classifications became too small for the GAN model to generate converged outputs. Therefore, we had to make a tradeoff and kept the ACUTMI and Normal databases with age and sex unsegregated. The age and sex distributions for the four categories of the ECG database are illustrated in Fig. 2. Unidentified sex is treated as Male and unknown age is treated as 50 by the DXL™ classifier.

*Setup and Training of the GAN*

We used TensorFlow 2.1 on an NVIDIA GeForce RTX 2080 Ti GPU as our platform and Keras API to train the 2D BiLSTM GAN. TensorFlow with Keras API provides convenient tools and interfaces for Python to build sequential generator and discriminator models. The generator model and discriminator model were built by adding a list of layers from a predefined stack. Bidirectional LSTM involves duplication of the first recurrent layer in the network, creating side-by-side layers, the first of which provides the input sequence without modifications while the second provides a reversed copy of the input sequence. The bidirectional data affords the model the opportunity to learn both from the past and from the future. The GAN comprises a generator and a discriminator. In the proposed architecture, the generator is taken to comprise a bidirectional LSTM layer and five 2D convolution layers, each with a kernel size of $16 \times 3$. The discriminator is taken to comprise four convolution layers, each with a kernel size of $16 \times 3$ as well, and a dense layer. The Leaky rectified linear unit (ReLU) activation layers were used between the convolutional layers in both the generator and the discriminator, which effectively prevented the vanishing gradient problem while simultaneously overcoming the "dying ReLU" issue induced by the negative inputs prevalent in ECG signals. Table 3 lists the parameters corresponding to each layer of the generator. Table 4 lists the parameters corresponding to each layer of the discriminator.

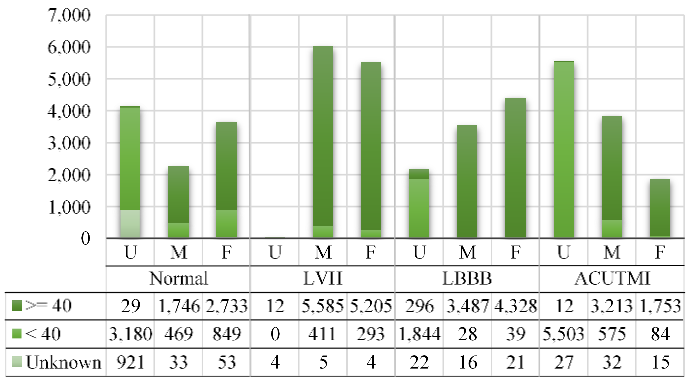

| | Normal | | | LVH | | | LBBB | | | ACUTMI | | |
|---|---|---|---|---|---|---|---|---|---|---|---|---|
| | U | M | F | U | M | F | U | M | F | U | M | F |
| >= 40 | 29 | 1,746 | 2,733 | 12 | 5,585 | 5,205 | 296 | 3,487 | 4,328 | 12 | 3,213 | 1,753 |
| < 40 | 3,180 | 469 | 849 | 0 | 411 | 293 | 1,844 | 28 | 39 | 5,503 | 575 | 84 |
| Unknown | 921 | 33 | 53 | 4 | 5 | 4 | 22 | 16 | 21 | 27 | 32 | 15 |

Fig. 2. The age and sex distributions for the four categories of the ECG database. M is male, F is female, U is Unidentified sex. U is treated as male and Unknown age is treated as 50 by the DXL™ classifier.

Table 3
Parameters Corresponding to Each Generator Layer

| Layer | Kernel size | Input size | Output size | # params |
|---|---|---|---|---|
| BiLSTM | N/A | 12 | 64 | 39,424 |
| Conv2d_1 | $16 \times 3$ | 16 | 128 | 98,432 |
| Conv2d_2 | $16 \times 3$ | 128 | 64 | 393,280 |
| Conv2d_3 | $16 \times 3$ | 64 | 32 | 98,336 |
| Conv2d_4 | $16 \times 3$ | 32 | 16 | 24,592 |
| Conv2d_5 | $16 \times 3$ | 16 | 1 | 769 |
| Total # parameters | | | | 654,833 |

Table 4
Parameters Corresponding to Each Discriminator Layer

| Layer | Kernel size | Input size | Output size | # params |
|---|---|---|---|---|
| Conv2d_1 | $16 \times 3$ | 1 | 32 | 1,568 |
| Conv2d_2 | $16 \times 3$ | 32 | 64 | 98,368 |
| Conv2d_3 | $16 \times 3$ | 64 | 128 | 393,344 |
| Conv2d_4 | $16 \times 3$ | 128 | 256 | 1,573,120 |
| Dense | N/A | 204,800 | 1 | 204,801 |
| Total # parameters | | | | 2,271,201 |

The input data used for training were taken to have a size of $400 \times 8$ (8 leads and 400 samples corresponding to each lead) for each of the ECGs in a batch of size 32. As seen in Table 1, for example for Normal category, we used 9,011 ECGs for training and 1,002 ECGs for subsequent plausibility check and statistical evaluation. The number of epochs required to obtain a synthetic ECG that is acceptable for further verification is determined empirically and depends on the defined criteria. With the help of visual inspections and a certain degree of trial and error, we set up a plausibility-check process to evaluate automatically the acceptance of a generated ECG that can be transmitted to the next step of the synthesis process without human intervention. The plausibility check included the following criteria: First, we measured the statistical distance (specifically, maximum mean discrepancies (MMDs)) between the generated ECG and all testing ECGs. The mean value of the MMDs was calculated and compared with a predefined threshold (0.004). An epoch with a generated ECG that had a mean MMD larger than this threshold was deemed to be unsuccessful. Consequently, the generated ECG was not transmitted to the next step, and the learning process was continued.

We observed that staying below the MMD threshold was not a sufficient condition for a desirable synthetic ECG: MMDs of ECGs may exhibit small values for reasons other than a good correspondence. For instance, a random distribution of small values could also result in small MMDs. Therefore, the second criterion of the plausibility check was whether the amplitude range of the generated ECG between the maximum and minimum amplitudes across all leads was observed to be smaller than a predefined threshold (1.2 mV): if so, the epoch was again deemed to be unsuccessful. The third criterion used for the plausibility check was whether the generated ECG exhibited erratic edges. Because our training data were regulated to have the QRS complex at the center of each ECG, the amplitude at both ends of the generated ECG signals should be relatively small—usually close to the isoelectric level. Therefore, if the absolute maximum amplitude of the generated ECG occurred at the edges, defined as the first or the last 50 samples, the epoch was deemed to be unsuccessful. The final criterion of the plausibility check was based on our observation that the passage of at least 10 epochs was required before the generated signals began to exhibit meaningful ECG features. (In most cases, 10 epochs were sufficient to induce the emergence of features, but the number depended on the ECG category. For instance, a greater number of epochs was required by the model to learn the Normal ECG features correctly. This is detailed in the Experimental Results section.) The thresholds to pass the plausibility check are summarized in Table 5.

Table 5
Thresholds to Pass Plausibility Check

| Max Mean MMDs | Min Amplitude Range (mV) | Erratic Edges | Min # of Epochs |
|---|---|---|---|
| 0.004 | 1.2 | Max does not occur in first or last 50 samples | 10 |

Two approaches can be followed after a generated ECG passes the plausibility check and the final verification:
- Relearning: Restart the training from scratch, i.e. epoch 0.
- Accumulative Learning: Continue the training for the next generated ECG based on the existing epochs.

Accumulative learning is more efficient at generating new ECGs, but there is a natural concern that the ECGs generated in adjacent epochs may be excessively similar to each other. Relearning, by contrast, is more likely to generate ECGs that differ from each other, because each round of training is initialized from a random distribution. However, it is computationally more expensive than accumulative learning.

In our experiments, we observed that, although the morphological characteristics of the generated ECGs improved statistically over the first hundred epochs, the model did not converge to a fixed data distribution (i.e., a fixed ECG profile) even after hundreds of epochs of training. (The analysis of the diversity of the ECGs synthesized via our method is presented in the Synthetic ECG Evaluation section.) Therefore, the theoretically possible lack of diversity in the generated ECGs does not provide sufficient justification to favor relearning over accumulative learning. This is one of the challenges faced by the GAN model, and is the reason that visual inspection is the method most commonly used to qualify a generated target and to decide whether sufficient learning has been achieved. We applied the same process to generate synthetic ECGs belonging to each of the four categories.

Fig. 3 depicts the losses for the generator and discriminator and the accuracy with which the discriminator distinguishes synthetic from real ACUTMI ECGs during a training round lasting 1000 epochs. It is evident that the losses corresponding to the discriminator decrease to 0.5 after 100 epochs and then fluctuate around 0.5 for the remainder of the 1000 epochs. The accuracy with which the discriminator distinguishes between synthetic and real ECGs increases to 0.95 after 100 epochs and then continues to fluctuate around that value for the remaining epochs.

### Post-Learning Data Processing

The ECGs generated during the model training that passed the plausibility check were automatically transmitted to the next stage, referred to as post-learning data processing. Each of the generated ECGs was represented by a $400 \times 8$ matrix, which contained 400 samples for each of the eight representative beats—one per lead. Leads I and II were used to augment the representative beats for the other four limb leads using (1). To generate a standard 12-lead 10-second ECG, we attached 13 of the same $400 \times 12$ representative beats to each other; of the resulting 10.4 second samples only the first 10 seconds were taken. As described previously, the RR intervals and rhythm information are not a part of the training dataset. Therefore, the synthetic RR intervals for the newly created 10-second ECGs were only meant to convert the synthetic ECG to the format expected by the DXL™ algorithm used during the verification stage. A realistic variation of RR intervals could have been introduced while the representative beats were being stitched together. In this study, however, we simply created ECGs with fixed RR intervals, because we were only interested in the morphology of synthetic ECG beats, not in the rhythm of 10-second ECGs. The newly created 10-second 12-lead ECGs were converted from numerical to XML format, as expected by the DXL™ algorithm.

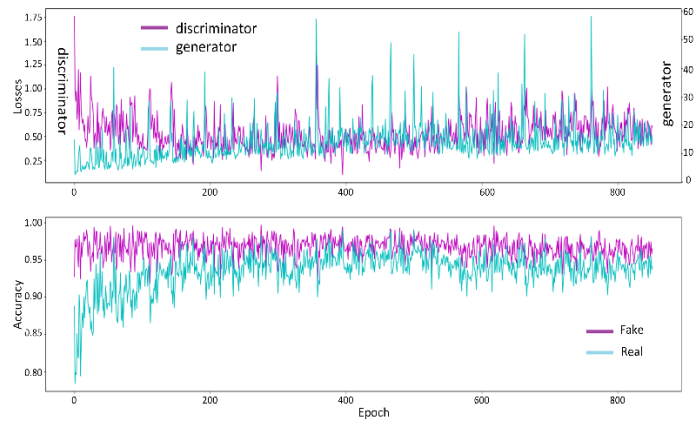

Fig. 3. Generator and discriminator losses (top) and discriminator accuracy (bottom) for 1000 training epochs for acute-myocardial-infarction ECGs.

### Verification of Synthetic 12-Lead ECGs

The synthetic 10-second 12-lead ECGs in the XML format were analyzed by the DXL™ algorithm during the final verification. As mentioned previously, for a given training cycle, the model was trained with ECGs from only one of the four categories—Normal, LVH, LBBB, or ACUTMI. The synthetic ECGs were verified for the same category. For instance, if a synthetic ECG was generated by a model trained with Normal ECGs, it was verified to be good if the DXL™ algorithm classified it as Normal. Otherwise, it was discarded and training continued with the next epoch. The same verification process was conducted for LVH, LBBB, and ACUTMI ECGs. The 12-lead ECGs that passed the final verification were the final products of the synthetic ECG generation process. As discussed previously, two alternative approaches can be followed after a synthetic ECG is marked as good. To generate the next synthetic ECG, the model can either restart from scratch (relearning), or continue to the next epoch (accumulative learning). The process workflow of ECG generation from training the model to final verification is depicted in Fig. 4.

## Experimental Results

We conducted 1000 epochs of training for the 2D BiLSTM GAN model to produce 10-second 12-lead ECGs for each of the four categories—Normal, LVH, LBBB, and ACUTMI. During the 1000 epochs of training, a certain percentage of generated

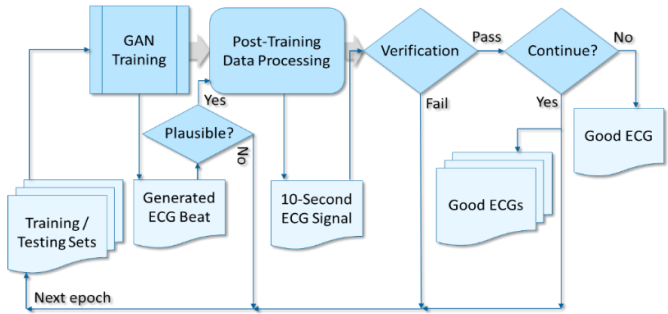

Fig. 4. Process workflow of ECG generation, from training the generative adversarial network (GAN) to final verification

ECGs failed the plausibility check. Consequently, the number of synthetic 12-lead ECGs transmitted to the final verification stage was less than 1000. The plausible synthetic 12-lead ECGs that passed the final verification were counted as *successes*. We defined the success rate of the synthetic ECG generation process to be the percentage of plausible ECGs that passed final verification. Table 6 lists the success rates corresponding to the four ECG categories.

Table 6
Success Rates of the Synthesis Process

| Category | *Plausible* | *Verified* | Success Rate (%) |
|---|---|---|---|
| Normal | 925 | 547 | 59.1 |
| LVH | 933 | 917 | 98.3 |
| LBBB | 927 | 865 | 93.3 |
| ACUTMI | 803 | 637 | 79.3 |

The success rate depended on the targeted morphology. As mentioned in the Data Preprocessing section, the training ECGs were not segregated by age and sex. When the DXL™ algorithm was used to classify the synthesized ECGs, it applied the default age and sex values of 50 and Male. This is not an issue for LBBB and LVH classifications. However, for ACUTMI some of the synthetic ECGs that are influenced by training ECGs from Female (V2/V3 ST elevation > 150 uV) may not pass the criteria for Male of age 50 (V2/V3 ST elevation > 200 uV). We believe this issue negatively affected the success rate for ACUTMI. As DXL™ is designed to rule out all possible abnormalities in an ECG before designating it as Normal, the success rate for Normal is lowest among those of the four categories.

Table 7 is the confusion matrix for the synthesized ECGs. Those (listed as Others) that did not fall into the four categories are further separated by their severities. The DXL™ algorithm assigns a severity level to each ECG in its diagnostic report (Table 8). For instance, a Normal ECG has a severity of Normal (NO). An otherwise normal ECG with a corrected QT interval shorter than 340 ms, a minor ST depression (< 0.03 mV) or elevation (< 0.05 mV), or some other minor condition has a severity code of Otherwise Normal (ON). The LVH, LBBB, and ACUTMI ECGs are associated with Abnormal severity (AB) except for a very small portion of LVH (< 10%) that could fall into Borderline (BO).

Table 7
The confusion matrix for the synthesized ECGs

| Actual <br> Predicted | Normal | LVH | LBBB | ACUTMI | Others (By Severity) | | | |
|---|---|---|---|---|---|---|---|---|
| | | | | | ON | BO | AB | DE |
| Normal | 547 | 53 | 0 | 1 | 118 | 79 | 127 | 0 |
| LVH | 0 | 917 | 1 | 0 | 0 | 0 | 14 | 1 |
| LBBB | 0 | 15 | 865 | 14 | 0 | 0 | 33 | 0 |
| ACUTMI | 0 | 20 | 5 | 637 | 0 | 4 | 136 | 1 |

Table 8
ECG Severity Levels Used by DXL™

| Severity | Code |
|---|---|
| Normal | NO |
| Otherwise Normal | ON |
| Borderline | BO |
| Abnormal | AB |
| Defective | DE |

As seen in Table 7, among the 378 synthesized ECGs incorrectly predicted to be Normal, 53 had morphology that deviated to LVH, 1 to ACUTMI, and 324 to Others (118 with severity ON, 79 with BO, and 127 with AB). The LVH had only 16 synthesized ECGs that failed to hit the target, with 1 deviating to LBBB, 14 to AB, and 1 to Defective (DE). The LBBB had 15 ECGs deviating to LVH, 14 to ACUTMI, and 33 to AB. The ACUTMI had 20 deviating to LVH, 5 to LBBB, 4 to BO, 136 to AB, and 1 to DE.

Fig. 5 is an example of a summary report generated by DXL™ for a successfully verified 12-lead synthetic ECG that mimics LBBB. Fig. 6 is a summary report for a 12-lead synthetic ECG targeted for LBBB but deviating to the other Abnormal (AB) morphological profile. Fig. 7 is a summary report for a 12-lead synthetic ECG targeted for Normal but deviating to Otherwise Normal because of minor ST depression.

## Synthetic ECG Evaluation

To investigate the characteristics of synthetic ECGs for data augmentation, we examined the similarities among generated ECGs, and those between them and the training set. The addition of synthetic ECGs very similar to each other or to existing data would not add much new information to an existing real database. Furthermore, if the profiles of synthetic ECGs are biased towards the training ECGs compared to the testing ECGs, the model may have been overfitted during training.

Our first approach involved six ECG features that are commonly used in clinical practice—PR Interval, QRS Duration, QT Interval, P-Axis, QRS-Axis, and T-Axis.

The second approach involved treating each ECG as a generic time series and measuring the statistical distances between the three different datasets—synthetic, training, and testing ECGs.

The results obtained via both approaches are presented as histograms. Three measurement parameters—the interquartile range (IQR), skewness, and kurtosis—together provide a quantitative insight into the shapes of the histograms and the diversity and variability of the measurement distributions. IQR is a measure of where the bulk (middle 50 %) of the values lie. Skewness is a measure of the lack of symmetry in the data. Kurtosis is a measure of whether the data are heavy- or light-tailed relative to a normal distribution. Fig. 8 illustrates how skewness and kurtosis are related to the shape of the histogram.

In the first approach, we used DXL™ to extract automatically the six features from the verified synthetic ECGs as well as from the training and testing ECGs. Because the size of the training dataset was much larger than that of the other two, we randomly selected a similar number of the training ECGs to ensure a fair comparison. As an example, Fig. 9 depicts the histograms for

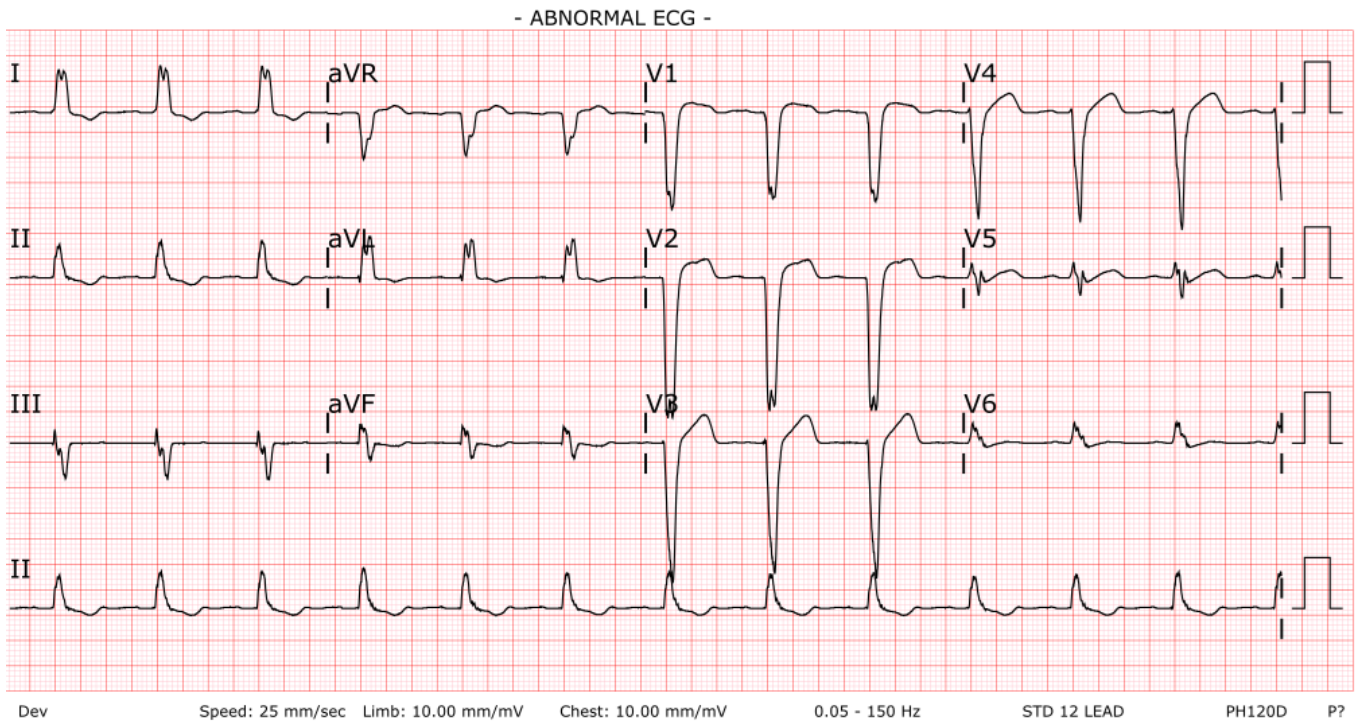

Fig. 5. An example of a summary report generated by DXL™ for a verified 12-lead synthetic ECG that mimics left bundle branch block.

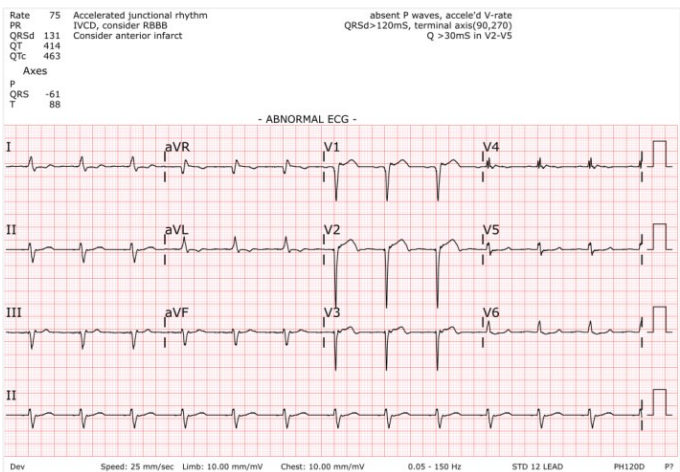

Fig. 6. A summary report for a 12-lead synthetic ECG targeting left bundle branch block but deviating to of other Abnormal (AB) morphological profile.

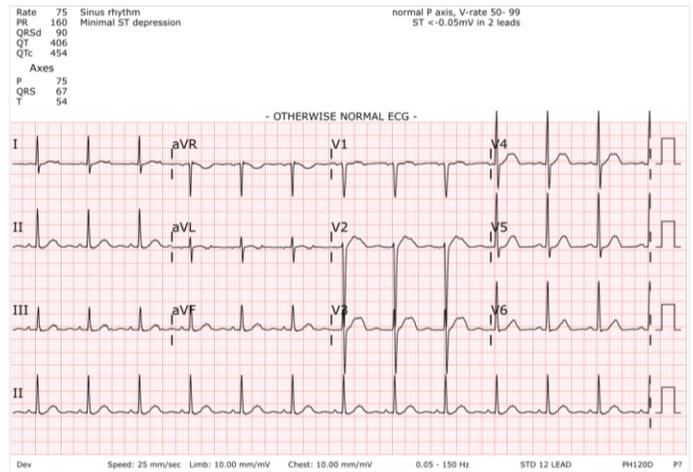

Fig. 7. A summary report for a 12-lead synthetic ECG targeting Normal but deviating to the Otherwise Normal (ON) morphological profile due to minor ST depression.

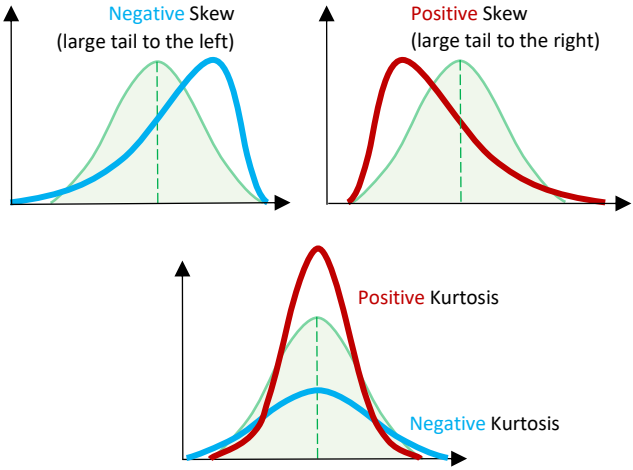

Fig. 8. Illustrations of skewness and kurtosis.

QRS durations in LVH ECGs. The fitted normal distributions and their mean values are also depicted. While the mean QRS duration for the synthetic ECGs appeared to be about 6 ms longer than for the training ECGs, we regarded that as a statistical fluctuation. The IQR, skewness, and kurtosis corresponding to each dataset are given in the legend box. QRS durations for the synthetic ECGs are observed to exhibit similar distributions to those of the training and testing ECGs. Further, there is no visible bias in the synthetic ECGs towards the training ECGs compared to that towards the testing ECGs. The quantitative metrics—IQR, skewness, and kurtosis—are also observed to be comparable in the three datasets. The bottom part of Fig. 9 is a scatter plot of skewness against mean QRS duration for all four categories in Table 6. (The LVH values are marked by stars.)

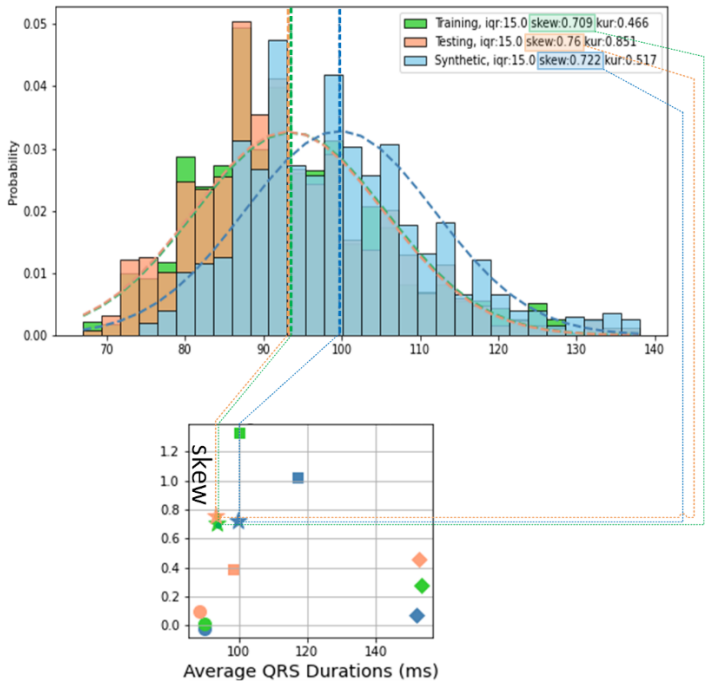

Fig.9. (Top) Histograms for QRS durations in synthetic (blue), training (green), and testing (orange) left-ventricular-hypertrophy (LVH) ECGs with corresponding interquartile range (IQR), skewness, and kurtosis. The mean values for the three histograms are shown by the vertical dashed lines. (Bottom) Stars indicate skew values from the histograms plotted against the histogram means. (Square, circular, and diamond symbols represent corresponding values for categories other than LVH: see Fig. 10.)

We measured the six ECG morphology features for Normal, LBBB, LVH, and ACUTMI ECGs. Fig. 10 depicts the IQR, skewness, and kurtosis values (Y-axis) corresponding to the means (X-axis), mapped from histograms in the same way as in the bottom of Fig. 9. Synthetic data points are highlighted in blue, training points in green, and testing points in orange. Normal data points are represented using circles, LBBB using diamonds, LVH using stars, and ACUTMI using squares.

As is evident from Fig. 10, no clear grouping is noticed for any particular color in most cases, i.e., there is no unique distribution for any of the three datasets. This, however, does not seem to apply to QT intervals and P-wave axis (although in lesser degree), the datapoints corresponding to which for the synthetic ECG appear to be segregated from the training and testing datapoints. This can be attributed to the fact that the ends of the T-waves of some training ECGs might have been truncated during the pre-training data preparation process. Consequently, the learning of the model is constrained to a shortened QT interval. The P-wave axis datapoints for the synthetic ECGs can be attributed to the delicate nature of P-wave morphology. They are usually small in value, exhibit low signal-to-noise ratios, and are prone to noise. Although the model is still able to learn the locations of the P-waves, as evidenced by the PR interval clusters (1st row), the shapes and directions of the P-waves seem to be more difficult for the model to learn.

Another observation from Fig. 10 is that the synthetic ECG datapoints (blue) are not any closer to the training ECG datapoints (green) than to the testing datapoints (orange), even in the cases of the QT intervals and the P-wave axis. In other words, the synthetic ECGs do not appear biased towards the training ECGs.

In the second approach, we used statistical distance measurements (SDMs)— percent root mean square difference (PRD) and root mean square error (RMSE)—to evaluate the similarities between the synthetic, training, and testing ECGs. PRD and RMSE are defined by (2) and (3), respectively, where N denotes the number of samples, and $x$ and $y$ denote the signals being compared:

$$PRD = \sqrt{\frac{\sum_{i=1}^{N}(x_i - y_i)^2}{\sum_{i=1}^{N}(x_i)^2}} \times 100 , \qquad (2)$$

$$RMSE = \sqrt{\frac{\sum_{i=1}^{N}(x_i - y_i)^2}{N}} . \qquad (3)$$

For each plausible synthetic ECG, we calculated SDMs against every ECG in the training set. This process generated, e.g., 9,011 (the size of the training set) SDMs for any given synthetic Normal ECG. Then, we identified the minimum, maximum, and mean SDMs. An identical match resulted in an SDM of zero. The final outcome of these calculations was a list of minimums, maximums, and means for RMSEs and PRDs between the synthetic and training ECGs. The size of each list was the same as the number of plausible synthetic ECGs. A similar process was repeated for synthetic versus testing ECGs, and testing versus training ECGs.

We then plotted the histograms of the SDM distributions for the three pairs of calculations along with the fitted normal distribution and their mean values. The measurements for variability and dispersion—IQR, skewness, and kurtosis—were also calculated. These measurements provide insight into the data distribution and the relationships between the different pairs. For example, Fig. 11 depicts the histograms of the minimum RMSE distributions for LBBB ECGs. The statistical distances between the synthetic and training ECGs are observed to lie between the synthetic-to-testing distances and the testing-to-training distances. In this case, the mean value of the RMSEs between the synthetic and training ECGs is close to, but a little smaller than, that of the RMSEs between synthetic and testing ECGs. However, the mean value of the RMSEs between the testing and training ECGs is observed to be even smaller. Therefore, it cannot be concluded in this case that the synthetic ECGs are biased towards the training ECGs. This is not surprising, as we did not identify any exact match between the synthetic and training ECGs (SDMs = 0).

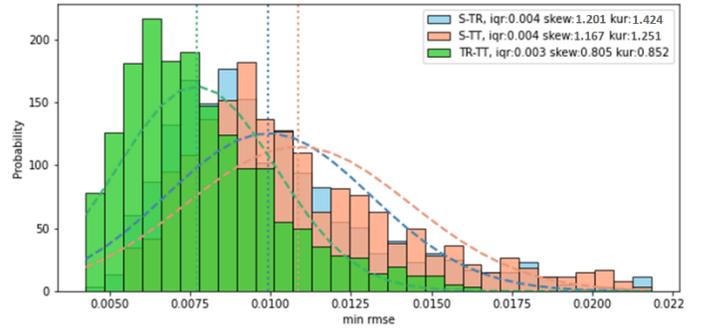

Fig. 11. Histograms of the minimum root mean square errors for left-bundle-branch-block ECGs measured between synthetic and training (blue), synthetic and testing (orange), and training and testing (green) ECGs. The mean values for the three histograms are shown by the vertical dashed lines.

As in the previous approach to measuring the six ECG features, we calculated the SDMs for Normal, LBBB, LVH, and ACUTMI ECGs. Fig. 12 depicts the IQR, skewness, and kurtosis (Y-axis) versus the corresponding means of the minimum, maximum, and mean SDMs (X-axis). The SDMs between synthetic and training ECGs are represented in blue, those between synthetic and testing ECGs in orange, and those between training and testing ECGs in green. Normal datapoints are represented using circles, LBBB points using diamonds, LVH points using stars, and ACUTMI points using squares.

As is evident from the figure, there is no clear clustering corresponding to any particular color; that is, there is no unique distribution for any of the three SDM datasets. In particular, there is no evidence to indicate that the SDMs between synthetic and training ECGs are much smaller than those between the other two datasets.

## Limitations

As only single representative beats—one per ECG lead—are used in our model, the RR intervals and rhythm information are not learned and generated by it. Therefore, it is only suitable for generating synthetic ECGs for those ECG categories that are primarily defined by the morphology of the average beat, but not by the rhythm. Models to generate abnormal ECG rhythms have already been extensively studied [37].

It is possible to teach the rhythm to the model by including a higher level of complexity in the ECG signals, e.g., by including multiple consecutive beats. Such an addition would demand a much larger training database, improved or new deep learning architectures, and a greater amount of computational power and resources.

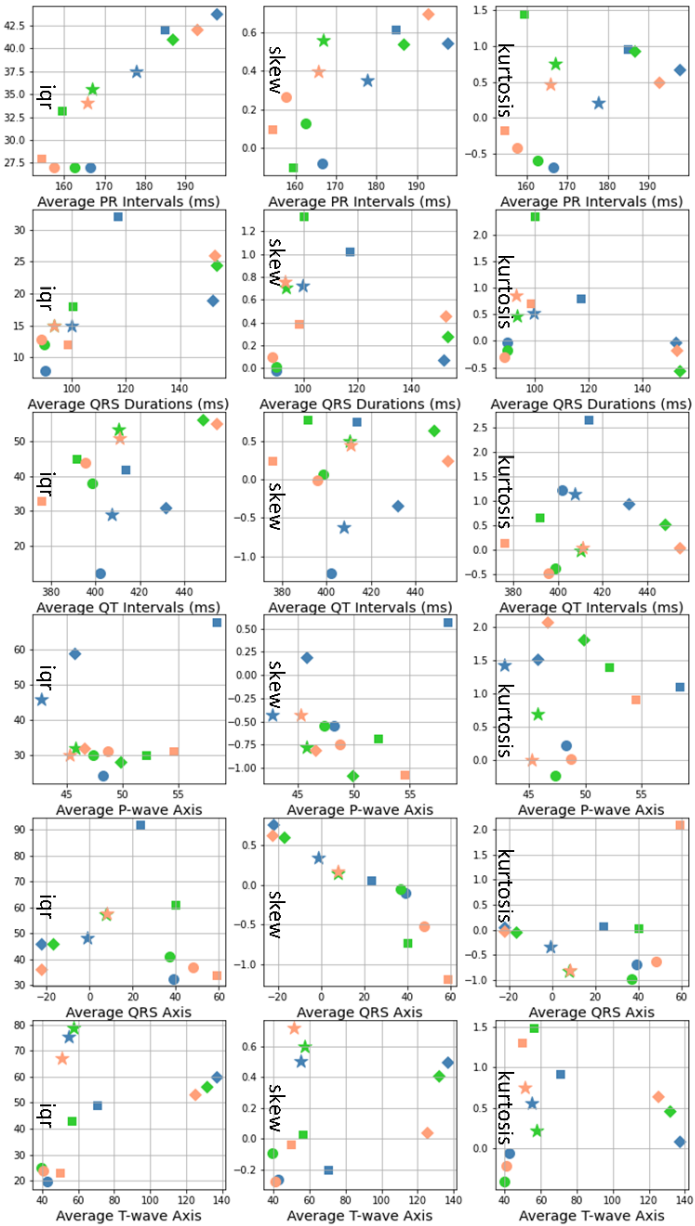

Fig. 10. Interquartile range (IQR), skewness, and kurtosis values corresponding to histogram means. Synthetic data points are highlighted in blue, training points in green, and testing points in orange. Normal data points are represented using circles, left bundle branch block using diamonds, left ventricular hypertrophy using stars, and acute myocardial infarction using squares.

## Conclusion

In this study, we proposed and trained a 2D BiLSTM GAN model to produce synthetic 12-lead ECGs belonging to four categories—Normal, LVH, LBBB, and ACUTMI—with acceptable success rates. We have demonstrated that the model was able to generate consistent standard 12-lead ECGs that are not only physiologically plausible but also can be verified to mimic the intended clinical conditions. The end-to-end ECG synthesis process is fully automatic and does not require any visual inspection. The characteristics of the synthetic 12-lead ECGs generated by our method were validated from three perspectives:

- The synthetic ECG features and morphological profiles demonstrate diversities and variabilities comparable to those of real ECGs;

- The synthetic ECGs are not biased towards the training dataset;

- They were not rejected by the commercial DXL™ diagnostic ECG algorithm as defective recordings.

The characteristics of the synthetic 12-lead ECGs strongly support the feasibility of using synthetic ECGs to augment databases of real ECGs possessing complex morphological features.

## Acknowledgements

We thank our colleague Richard Gregg for his assistance with the DXL™ Algorithm.

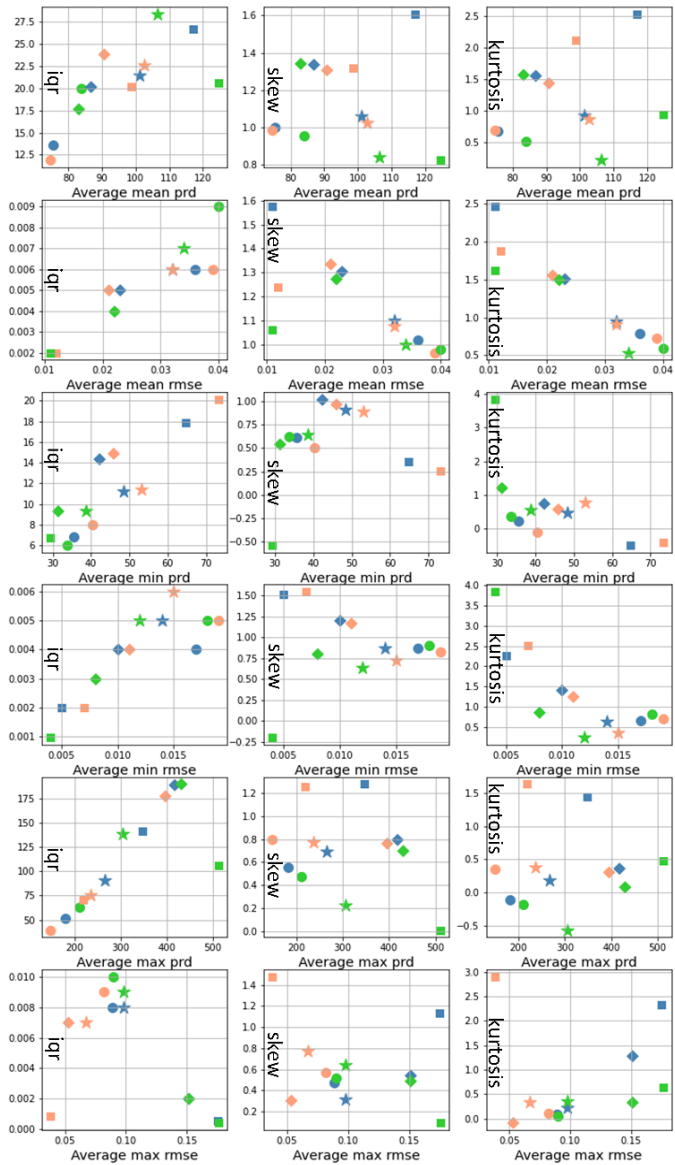

Fig. 12. Interquartile range (IQR), skewness, and kurtosis values corresponding to the means of the minimum, maximum, and mean statistical distance measurements (SDMs). The SDMs between synthetic and training ECGs are represented in blue, those between synthetic and testing ECGs in orange, and those between training and testing ECGs in green. Normal data points are represented using circles, left bundle branch block using diamonds, left ventricular hypertrophy using stars, and acute myocardial infarction using squares.